\documentclass[aps,prc,floatfix,showpacs]{revtex4}

\usepackage{graphicx}
\usepackage{amsmath}

\begin{document}
%\preprint{DUKE-TH-03-242}

\title{The Decoherence Time in High Energy Heavy Ion Collisions}
\author{Berndt M\"uller}
\affiliation{Department of Physics, Duke University, 
             Durham, North Carolina 27708} 
\author{Andreas Sch\"afer}
\affiliation{Institut f\"ur Theoretische Physik,
             Universit\"at Regensburg,
             D-93040 Regensburg, Germany}
\date{\today}

\begin{abstract}
We calculate the decoherence time of the ground state wave function of 
a nucleus in a high energy heavy ion collision. We define this time as 
the decay time of the ratio ${\rm Tr} D^2/({\rm Tr} D)^2$ of traces of 
the density matrix $D$. We find that this time is smaller or equal to 
$1/Q_s$, where the saturation scale $Q_s$ is defined within the color 
glass condensate model of parton saturation. Our result supports the
notion that the extremely rapid entropy production deduced for the 
early stage of heavy ion collisions at collider energies is to a large 
extent caused by the decoherence of the initial-state wave functions.
\end{abstract}
\maketitle
\section{I. Introduction}
The physics program of the Relativistic Heavy Ion Collider (RHIC) has 
produced many intriguing results and posed a number of unexplained problems.
One central question that has emerges is: How can hydrodynamical 
behaviour, implying local thermal equilibration and complete decoherence 
of the initial state, occur on a time scale which is considerably shorter 
than 1 fm/$c$? \cite{rev1,rev2}.

In an earlier article \cite{ba1} we showed that the entropy per rapidity 
interval produced by decoherence alone is proportional to $(RQ_s)^2$,
where $R$ is the nuclear radius and $Q_s$ is the gluon saturation scale
\cite{Qs-ref}. The resulting entropy per unit rapidity interval is of 
the order $1000-2000$, which amount to a substantial fraction of the 
total produced entropy. We than argued that the decoherence time has 
to be of order $1/Q_s$, as this is the natural scale of the process.
The purpose of the present article is to substantiate this claim by 
means of a quantitative calculation.

The fact that entropy produced by decoherence can play an important role
in high-energy heavy-ion collisions was to the best of our knowledge first discussed 
in \cite{elze}.  

Our article is organized as follows: In Section II we explain how we 
attack the problem and how we define the decoherence time. In Section 
III we describe the calculation of the density matrix for gluons, which 
undergo a hard scattering process, in detail. Our calculation is based 
on work by Kovchegov and Mueller \cite{Yuri1,Yuri2} on gluon production 
in heavy ion collisions. We generalize their derivation of probabilities 
to the level of density matrices. In Section IV we calculate the 
decoherence time using the results from Section III.

\section{Strategy}

Our goal is to obtain an estimate of the decoherence time of the
gluon distribution in a large nucleus (1), when it is hit by another
very energetic large nucleus (2). In principle, the decoherence 
process is encoded in the time evolution of the density matrix 
$\hat D$ in a very simple manner: For vanishing off-diagonal 
matrix elements the system is completely decoherent. In our case 
the dominant degrees of freedom are the gluons and thus the relevant 
density matrix is that of the gluons in nucleus 1. Decoherence thus 
manifests itself in a gradual disappearance of its off-diagonal 
elements $D_{k_1,k'_1}$ with the momenta $k_1\neq k'_1$. 

This sounds simple enough, but actually calculating the time 
evolution of off-diagonal elements in a complex multi-particle 
state is an extremely  difficult task \cite{deco1,deco2}. 
Therefore, one has studied so far mainly very simple toy models, 
like one harmonic oscillator in a termal bath of other harmonic 
oscillators, or rather specific situations, like neutrino oscillations 
\cite{deco3}.

Luckily, also the situation encountered in high energy heavy ion 
collisions is such a special case, for the following two reasons:\\ 
i) All gluons which undergo scattering are boosted into a part of phase space which was originally 
empty.\\
ii) The description simplifies strongly in the rest system of one of the 
colliding nuclei. For an arbitrary Lorentz frame we do not know how to model
the degree of coherence before the collision in both nuclei. In the rest frame of 
one of the nuclei, however, 
both nuclei can be approximated by different  asymptotic descriptions. 
For a very fast moving nucleus one observes saturation and can describe the 
gluon field correlations along the lines of \cite{Yuri1,Yuri2}, while 
for the nucleus at rest we can assume a simple {\em isotropic}
Gaussian correlation of gluons with a typical virtuality of order $1/\lambda$.
The precise form of this correlation does not really matter, because 
we will only study the correlations in that part of phase space mentionned in i),
which is originally empty.\\
Therefore, the elementary process of hard gluon scattering
is sufficent to estimate the decoherence time.\\

\begin{figure}[tb]   
\resizebox{0.8\linewidth}{!}
          {\rotatebox{0}{\includegraphics{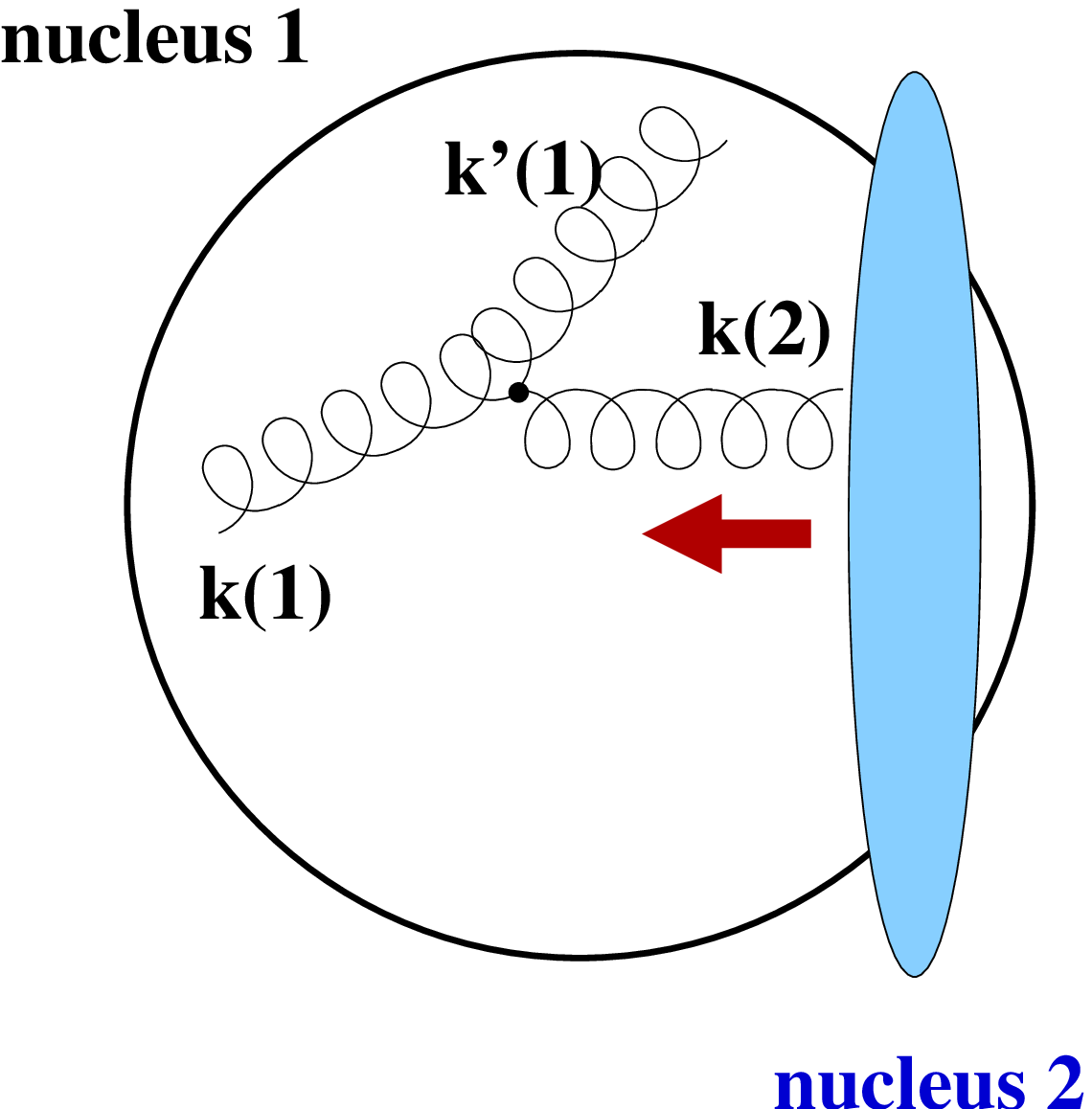}}\hskip 6 cm
          {\includegraphics{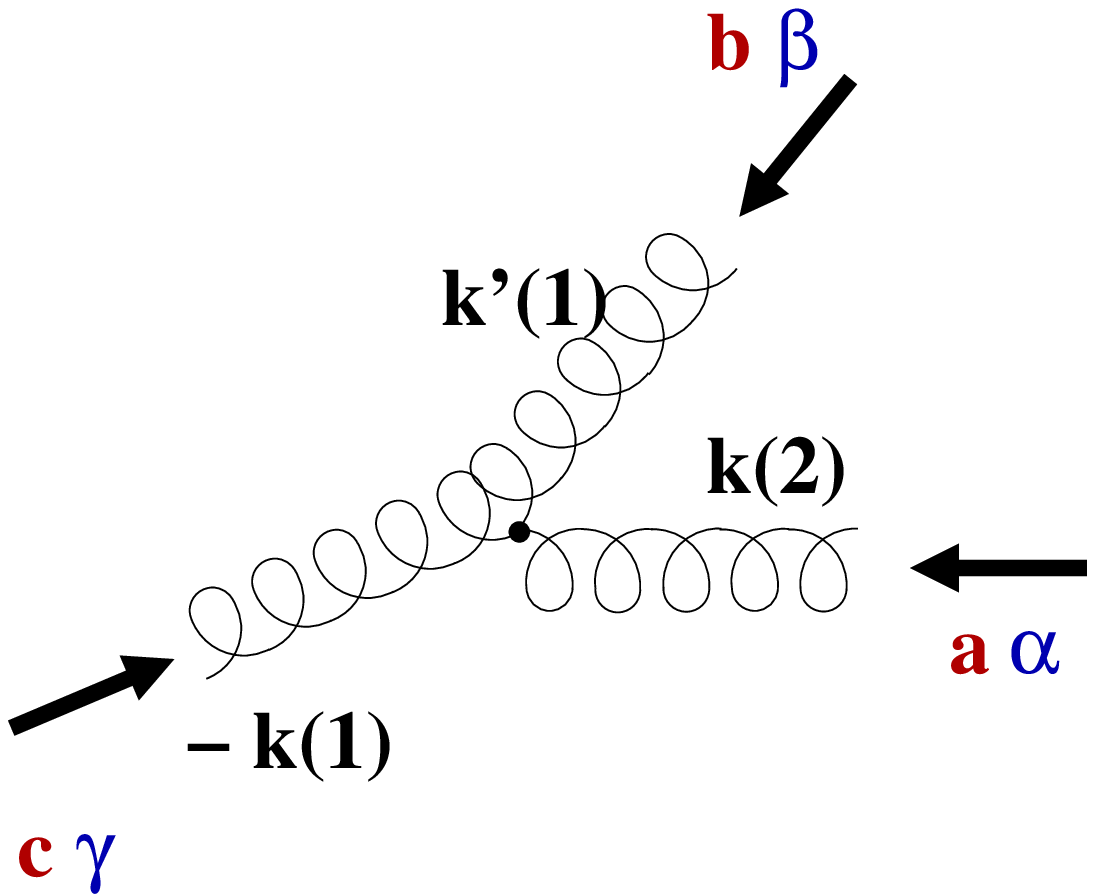}}}
\caption{The chosen frame and notation}
\label{fig1}
\end{figure}

We perform our calculation in the rest frame of 
nucleus 1, which is hit by the highly Lorentz contracted nucleus 2, 
which is moving to the left. At lowest order in the strong coupling
$\alpha_s$ the time evolution of the gluon density matrix is 
governed by the reaction $G_1+G_2\to G'_1$, where $G_i$
denotes a component of the gluon wave function of nucleus $i$ 
(see Figure \ref{fig1}).  

We investigate the gluon density matrix $\hat D(t)$ in a plane wave 
basis. Because the ground state wave function of nucleus 1 is a 
fully coherent bound state, the initial density matrix $\hat D(0)$ 
is strongly non-diagonal in this basis. As usual for a fast moving 
projectile, it is convenient to use light-cone coordinates 
$(x_+,x_-,x_{\perp})$ with $x^{\pm} = (x^0 \ pm x^3)/\sqrt(2)$.
Since, in our convention, nucleus 2 moves to the left,
$k_2^- = (k^0-k^3)/\sqrt(2) = k_{2+}$ is large and the nucleus 
remains localized in $x^+=x_-$, which thus denotes the light-cone
position of nucleus 2. On the other hand, $x^-=x_+ \approx \sqrt{2}x^0$
takes on the role of the ``light-cone time'', by which the progress
of the collision is monitored. This is also the starting idea of the 
Color Glass Condensate (CGC) approach \cite{weigert}, which is based on
the assumption that most of the gluons in nucleus 2 hardly evolve on the 
time scale of the collision, endowing the gluon distribution with a 
``glassy'' nature.

The ``time'' evolution of $\hat D$ due to the three-gluon interaction 
is then given by
\begin{eqnarray}
\hat D(dx_+) &=& \hat U(dx_+) \hat D(0) \hat U ^{\dagger}(dx_+) 
\nonumber \\
&=& \hat D(0) - {\rm i} [\hat H_{\rm int},\hat D(0)] \frac{dx_+}{\sqrt{2}}
    - [\hat H_{\rm int},[\hat H_{\rm int},\hat D(0)]]\frac{(dx_+)^2}{4} 
    + \ldots
\label{eq1}
\end{eqnarray}
with
\begin{equation}
\hat U(dx_+) = \exp(-{\rm i}\hat H_{\rm int}dx_+/\sqrt{2}) .
\label{eq2}
\end{equation}

For what follows, the variable $r$ stands for the multiple quantum 
numbers completely specifying a gluonic state in nucleus 1:
$r=\{k,\epsilon^\beta,b\}$, with wave vector $k$, polarization
vector $\epsilon^\beta$, and color index $b$. The matrix elements 
of $H_{\rm int}$ between two plane wave gluon states of nucleus 1 
are then given by
\begin{eqnarray}
H_{r_1,r'_1} &\equiv & \langle r_1 | H_{\rm int} | r'_1 \rangle 
= \int 
d^2x_{\perp}dx_- \int \frac{dk_{2+}d^2k_{2\perp}}{(2\pi)^3}
e^{{\rm i}(k_{2+}x_- -k_{2\perp}\cdot x_{\perp})} 
 e^{{\rm i}[(k_{1+}-k'_{1+})x_- -(k_{1\perp}-k'_{1\perp})\cdot x_{\perp}]} 
\nonumber \\
&& \qquad 
gf_{abc} \left[ g_{\alpha\beta}\left(k_2-k_1\right)_{\gamma} 
               +g_{\beta\gamma}\left(k_1+k'_1\right)_{\alpha} 
               +g_{\gamma\alpha}\left(-k'_1-k_2\right)_{\beta} \right] 
\nonumber \\
&& \qquad \qquad
  [2k_{1+}k'_{1+}V^2]^{-1/2} \epsilon_1^{\beta} {\epsilon'}_1^{\gamma}  
  2 {\rm Tr} \left[ T^a A_{\perp}^{\alpha}(k_{2\perp},k_{2+})\right] .
\label{eq3}
\end{eqnarray}
Here $1/\sqrt{\sqrt{2}k_{1+}V}$ is the normalization factor 
for a plane wave gluon in the light-cone formalism. $V$ is some 
appropriate $(-,\perp)$ normalization volume, which we will specify later.

$H_{r_1,r'_1}$ describes the interaction with the incident gluons
contained in nucleus 2, which moves very close to the speed of light.
Since these gluons originate in many different nucleons, it is a
good approximation to consider them as uncorrelated and to describe 
them as a Gaussian random ensemble \cite{weigert}
representing an incident stream of Weizs\"acker-Williams (WW) gluons, 
all of which have positive $k_{2+}$ momentum.

The gluon fields of 
nucleus 1 will be characterized by a Gaussian ensemble of off-shell
gluon fields with a spatial coherence length $\lambda$ of the order of
the confinement scale $\Lambda_{\rm QCD}^{-1}$:
\begin{equation}
\langle A^a_-(x) A^b_-(x') \rangle
= C_{ab} exp\left\{ -[(x'_{\perp} - x_{\perp})^2 + (x'_- - x_-)^2
  + (x'_+ - x_+)^2]/\lambda^2 \right\}  .
\label{eq4a}
\end{equation}
We emphasize that we do not assume the gluon field in any nucleus to be
dilute. We only assume that the typical momentum transfere $k_{\perp}$ is
sufficiently large so that we can treat the interaction perturbatively
in the coupling constant $\alpha_s=g^2/4\pi$.

The WW gluon fields are given in \cite{Yuri2}, eqs.~(1--3). For our 
calculation it is crucially important that the color charge densities  
$\rho^a(x_{\perp},x_-)$ and path ordered factors
\begin{equation}
S_0(x_{\perp},x_-) \equiv {\cal P} \exp\left( igT^a \int d^2z_{\perp}
   \Theta(z_- -x_-) \hat \rho^a(z_{\perp},z_-) 
   \ln \left( \left| x_{\perp}-z_{\perp} \right| \mu \right)\right)
\label{eq4}
\end{equation}
are uncorrelated for different values of $x_-$:
\begin{equation}
\langle \hat\rho^a(x_{\perp},x_-) \hat\rho^b(z_{\perp},z_-)\rangle
   = \frac{\alpha_s}{2N_c\pi} \rho_N(x_{\perp},x_-) 
     \delta(x_- -z_-)\delta^{ab} \delta^2(x_{\perp}-z_{\perp}) .
\label{eq5}
\end{equation}
This and eq.~(48) of \cite{Yuri1} allows to perform the calculation.

Let us stress again that the asymptotic descriptions (\ref{eq4a})
and eq.~(48) of \cite{Yuri1} treat the two nuclei in asymmetrically.
Therefore, one cannot expect the final expressions to be manifestly
boost invariant. Our calculation of the decoherence is specific to our
selected Lorentz frame. While a manifestly boost invariant treatment of
the decoherence process would be desirable, such a treatment would have
to rely on a Lorentz covariant representation of the nuclear ground
state. We do not address this interesting problem here.

Because the incident nucleus 2 is a color singlet, its glue field
vanishes on average: $\langle A_{\perp}^{\alpha}(k_2)\rangle =0$. 
This implies the absence of a contribution to the time evolution
of the diagonal elements of the density matrix $\hat D$ in first
order in $g$. The first nonvanishing term is thus of order $g^2$,
corresponding to the term proportional to $(x_+)^2$ in (\ref{eq1}),
which is of second order in the WW fields of nucleus 2.

The leading term for the evolution of the 
density matrix arises from the last term in eq.~(\ref{eq1}):
\begin{eqnarray}
D_{r,\hat r}(\tau) &=& D_{r,r}(0) + 
\sum_{r',\tilde r} 
W_{r,r'}(\tau) D_{r',\tilde r}(0) W_{\hat r,\tilde r}^*(\tau) .
\label{eq6}
\end{eqnarray}
where
\begin{equation}
W_{r_1,r'_1}(\tau) = \int_{-\tau/2}^{\tau/2} dx_+ H_{r_1,r'_1}(x_+) .
\label{eq7}
\end{equation}

Our aim is to calculate the density matrix in the region of phase space
populated by the outgoing gluons (momentum $k(1)$). This density matrix 
contains all the crucial information:
\begin{itemize}

\item 
The ratio
\begin{equation}
\frac{{\rm Tr}~\{D^2\}}{({\rm Tr}~\{D\})^2} 
\label{eqTrace}
\end{equation}
is a measure for the coherence. For a pure state it is one, for a 
completely decoherent state it is much smaller than one, of the order of 
$1/N$ for a region of phase space with $N$ states. We shall show 
that in the transverse directions (i.e. except for $k_+$)  we get nearly complete 
decoherence. We will also see that this ratio depends on the observation time 
and we shall define as decoherence time $\tau_{\rm deco}$ as the time after which it reaches 
$1/e$.

\item
The entropy of the final state can be calculated from 
\begin{equation}
S(\tau_f) = {\rm Tr}~\{D(\tau_f) \log D(\tau_f)\} .
\label{entropy}
\end{equation}
We argued in \cite{ba1} that this is a sizeable fraction of the total entropy 
produced in a heavy ion collision. It should be possible to  
evaluate Eq.(\ref{entropy}) 
for the density matrix we obtain, but the calculation is highly non-trivial.
Therefore, we leave this task for a future publication. 
Having a model for the density matrix 
in hand it should actually also be possible to calculate other quantities specifying
the initial state for the time evolution of the high-temperature phase produced in such collisions.\\

\end{itemize}

The symbol $\langle \cdots \rangle$ indicates and average over
the WW fields of nucleus 2.

\section{The density matrix resulting from hard gluonic interactions}

For the WW fields of nucleus 2 we have $k_{2-}=0$, 
because the fields $A(x)$ do not depend on $x_+$.
We therefore get
\begin{eqnarray}
\langle A^{\alpha'}(k'_2) A^{\tilde\alpha}(\tilde k_2)\rangle
&=& \int d^4x' d^4{\tilde x} e^{-ik'_2 x' + i{\tilde k}_2 \tilde x}
    \langle A^{\alpha'}(x') A^{\tilde\alpha}(\tilde x) \rangle
\\
&=& (2\pi)^2 \delta(k'_{2-})\delta(\tilde k_{2-})
    \int d^2x'_{\perp} d^2{\tilde x}_{\perp} dx'_- d{\tilde x}_- 
    e^{-{\rm i}(k'_{2+} x'_- -{\tilde k}_{2+}{\tilde x}_-
       -k'_{2\perp} x'_\perp +{\tilde k}_{2\perp}{\tilde x}_\perp)}
    \langle A^{\alpha'}(x'_\perp,x'_-) 
            A^{\tilde\alpha}(\tilde x_\perp,\tilde x_-) \rangle .
\nonumber 
\label{eq9}
\end{eqnarray}
The relevant term is now:
\begin{eqnarray} 
{\cal W}_{1\hat 1,1'\tilde 1} &\equiv& W_{r_1,r'_1} W_{\hat r_1,\tilde r_1}^* 
\nonumber \\
&=& 
\int dk'_{2-} \int d \tilde k_{2-} \delta(k'_{2-}+k'_{1-}-k_{1-})
  \delta({\tilde k}_{2-}+{\tilde k}_{1-}-\hat k_{1-})  
\nonumber \\
&&  \int dk'_{2+}d^2k'_{2\perp} \theta(k'_{2+}) 
    \int d{\tilde k}_{2+}d^2{\tilde k}_{2\perp} 
\theta({\tilde k}_{2+}) 
    \delta(k'_{2+}+k'_{1+}-k_{1+})
    \delta^2(k'_{2\perp}+k'_{1\perp}-k_{1\perp})
\nonumber \\
&&  \delta({\tilde k}_{2+}+{\tilde k}_{1+}-\hat k_{1+})
    \delta^2({\tilde k}_{2\perp}+{\tilde k}_{1\perp}-\hat k_{1\perp})
\nonumber \\
&& gf_{abc} \left[ g_{\alpha\beta}\left(k'_2-k'_1\right)_{\gamma} 
  +g_{\beta\gamma}\left(k'_1+k_1\right)_{\alpha} 
  +g_{\gamma\alpha}\left(-k_1-k'_2\right)_{\beta} \right] 
\nonumber \\
&& gf_{\tilde a\tilde bc} \left[ g_{\tilde\alpha\tilde\beta}
    \left(\tilde k_2-\tilde k_1\right)_{\tilde\gamma} 
  +g_{\tilde\beta\tilde\gamma}\left(\tilde k_1+\hat k_1\right)_{\tilde\alpha} 
  +g_{\tilde\gamma\tilde\alpha}\left(-\hat k_1-\tilde k_2\right)_{\beta} \right] 
\nonumber \\
&& 
[2 k_{1+} \hat k_{1+} V^2]^{-1/2}
   \epsilon(k'_1)^{\beta} \epsilon^*(\tilde k_1)^{\tilde\beta}  
\Bigl( \sum_{\epsilon}  \epsilon(k_1)^{\gamma} \epsilon^*(\hat k_1)^{\tilde\gamma}  \Bigr)
\nonumber \\ 
&& 2{\rm Tr}\Big[T^a A_{\perp}^{\alpha}(k'_{2\perp},k'_{2+})\Big]  
   2{\rm Tr}\Big[T^{\tilde a} A_{\perp}^{\tilde\alpha*}
    (\tilde k_{2\perp},\tilde k_{2+})\Big] .
\label{eq10}
\end{eqnarray}
Here the first line (the $k_{2-}$ integrals serves more or less only as 
reminder.)
Note that there are no normalization factors for the incomming gluon 
states, i.e. no factor $[2k'_{1+}\tilde k_{1+}V^2]^{-1/2}$. The reason is that 
our initial states are highly virtual, interfering  gluon states, i.e. no 
momentum eigenstates.                
For the time being 
we absorb all normalization factors into our definition of $d(x,x')$ which 
therefore is not dimensionles but  in coordinate space has the dimensions 
Energy$^{2}$. We do not know how to properly normalize a density matrix for an 
arbitrary virtual field configuration. Luckily we will only need the 
normalization of the diagonal elements.\\

We first focus on the ensemble average of the expression in the last
line, involving the WW fields in nucleus 2:
\begin{eqnarray} 
\langle {\cal J}\rangle &\equiv& V^{-2} \left\langle 
   2{\rm Tr}\Big[T^a A_{\perp}^{\alpha}(k'_{2\perp},k'_{2+})\Big]  
   2{\rm Tr}\Big[T^{\tilde a} A_{\perp}^{\tilde\alpha*}
    (\tilde k_{2\perp},\tilde k_{2+})\Big] \right\rangle 
\nonumber \\
&=& V^{-2} \int d^2w_{\perp}dw_-
\delta(k'_{2-}) \delta(\tilde k_{2-})
    e^{-{\rm i}(k'_{2+}w_- -k'_{2\perp}\cdot w_{\perp})} 
    \int d^2{\tilde w}_{\perp}d{\tilde w}_- 
    e^{{\rm i}({\tilde k}_{2+}{\tilde w}_- 
       -{\tilde k}_{2\perp}\cdot{\tilde w}_{\perp})} 
\nonumber \\
&& \int d^2y_{\perp}dy_- \theta(y_- -w_-)
   \int d^2{\tilde y}_{\perp}d{\tilde y}_- 
        \theta({\tilde y}_- -{\tilde w}_-)
   \frac{(w_{\perp}-y_{\perp})^{\alpha}}{|w_{\perp}-y_{\perp}|^2}
   \frac{(\tilde w_{\perp}-\tilde y_{\perp})^{\tilde \alpha}}
        {|\tilde w_{\perp}-\tilde y_{\perp}|^2}
\nonumber \\
&& \left\langle 2{\rm Tr} 
   \Big[T^a S_0(w_{\perp},y_-) T^d S_0^{-1}(w_{\perp},y_-)\Big] 
   2{\rm Tr}\Big[T^{\tilde a} S_0(\tilde w_{\perp},\tilde y_-) 
    T^{\tilde d} S_0^{-1}(\tilde w_{\perp},\tilde y_-) \Big] 
   \hat\rho^d(y_{\perp},y_-) 
   \hat \rho^{\tilde d}(\tilde y_{\perp},\tilde y_-) \right\rangle .
\label{eq14}
\end{eqnarray}

We now assume complete factorization of the average, i.e.
\begin{eqnarray} 
&& \left\langle 
   2{\rm Tr} \Big[T^a S_0(w_{\perp},y_-) T^d S_0^{-1}(w_{\perp},y_-)\Big] 
   2{\rm Tr} \Big[T^{\tilde a} S_0(\tilde w_{\perp},\tilde y_-) T^{\tilde d}
          S_0^{-1}(\tilde w_{\perp},\tilde y_-) \Big] 
\hat\rho^d(y_{\perp},y_-) \hat\rho^{\tilde d}(\tilde y_{\perp},\tilde y_-) 
\right\rangle
\nonumber \\
&=&
\left\langle 
2{\rm Tr} \Big[T^a S_0(w_{\perp},y_-) T^d S_0^{-1}(w_{\perp},y_-)\Big] 
2{\rm Tr} \Big[T^{\tilde a} S_0(\tilde w_{\perp},\tilde y_-) T^{\tilde d}
           S_0^{-1}(\tilde w_{\perp},\tilde y_-) \Big] \right\rangle
\langle \hat\rho^d(y_{\perp},y_-) 
        \hat\rho^{\tilde d}(\tilde y_{\perp},\tilde y_-) \rangle
\nonumber \\
&=&
\left\langle 
2{\rm Tr} \Big[T^a S_0(w_{\perp},y_-) T^d S_0^{-1}(w_{\perp},y_-)\Big] 
2{\rm Tr} \Big[T^{\tilde a} S_0(\tilde w_{\perp},\tilde y_-) T^{\tilde d}
          S_0^{-1}(\tilde w_{\perp},\tilde y_-) \Big] \right\rangle
\nonumber \\
&&
\times \frac{\alpha_s}{2\pi N_c}\rho_N(y_{\perp},y_-) \delta^{d \tilde d}
  \delta(y_- -\tilde y_-) \delta^2(y_{\perp}-\tilde y_{\perp}) .
\label{eq12}
\end{eqnarray}
Using the identity
\begin{equation}
2{\rm Tr} \Big[T^d B^e T^e \Big] 
2{\rm Tr} \Big[T^d C^f T^f \Big]=B^d C^d=  
2{\rm Tr} \Big[B^e T^e C^f T^f\Big]
\label{eq13}
\end{equation}
to rewrite the product of two color traces as a single trace, we get 
\begin{eqnarray} 
\langle {\cal J}\rangle &=& V^{-2} \int d^2w_{\perp}dw_- 
  e^{-{\rm i}(k'_{2+}w_- -k'_{2\perp}\cdot w_{\perp})} 
  \int d^2\tilde w_{\perp}d\tilde w_- 
  e^{{\rm i}({\tilde k}_{2+}{\tilde w}_- -{\tilde k}_{2\perp}
     \cdot{\tilde w}_{\perp})} \delta(k'_{2-}) \delta(\tilde k_{2-})
\nonumber \\
&&
\int d^2y_{\perp}dy_- \theta(y_- -w_-) \theta(y_- -{\tilde w}_-)
  \frac{(w_{\perp}-y_{\perp})^{\alpha}} {|w_{\perp}-y_{\perp}|^2}
  \frac{({\tilde w}_{\perp}-y_{\perp})^{\tilde\alpha}}
       {|{\tilde w}_{\perp}-y_{\perp}|^2}
  \frac{\alpha_s}{2\pi N_c}\rho_N(y_{\perp},y_-) 
\nonumber \\
&& \left\langle 
   2{\rm Tr} \Big[S_0^{-1}(w_{\perp},y_-) T^a S_0(w_{\perp},y_-) 
   S_0^{-1}({\tilde w}_{\perp},y_-) T^{\tilde a} 
   S_0({\tilde w}_{\perp},y_-) \Big] \right\rangle .
\label{eq14a}
\end{eqnarray}
Now we use eq.~(47) from \cite{Yuri1} and assume that for 
$a\neq \tilde a$ the ensemble average vanishes. The fact that the
original expression (47) has $S_0$ and $S_0^{-1}$ factors interchanged
and thus is the complex conjugate of our expression, does not matter,
because the result is real.
\begin{eqnarray}
&& \left\langle
   {\rm Tr} \Big[S_0^{-1}(w_{\perp},y_-) T^a S_0(w_{\perp},y_-) 
   S_0^{-1}({\tilde w}_{\perp},y_-) T^{\tilde a} 
   S_0({\tilde w}_{\perp},y_-) \Big] \right\rangle
\nonumber \\
&=& \frac{\delta^{a \tilde a}}{N_c^2-1} \left\langle
    {\rm Tr} \Big[S_0^{-1}(w_{\perp},y_-) T^a S_0(w_{\perp},y_-) 
    S_0^{-1}({\tilde w}_{\perp},y_-) T^{a} 
    S_0({\tilde w}_{\perp},y_-) \Big] \right\rangle
\nonumber \\
&=& \frac{\delta^{a \tilde a}C_F N_c}{N_c^2-1} 
    \exp \left(-g^2\frac{\pi\rho_{\rm rel}N_c}{4(N_c^2-1)} 
      xG(x,|\tilde w_{\perp}-w_{\perp}|^{-2})
      ({\tilde w}_{\perp}-w_{\perp})^2(y_- +y_-^{(0)}) \right)
\label{eq15}
\end{eqnarray}
where $y_-^{(0)} = \sqrt{R^2-x_{\perp}^2}/\sqrt{2}\gamma$.

This gives the following result:
\begin{eqnarray} 
\langle {\cal J}\rangle &=& V^{-2} \int d^2w_{\perp}dw_- 
  e^{-{\rm i}(k'_{2+}w_- -k'_{2\perp}\cdot w_{\perp})} 
  \int d^2\tilde w_{\perp}d\tilde w_- 
  e^{{\rm i}({\tilde k}_{2+}{\tilde w}_- -{\tilde k}_{2\perp}
     \cdot {\tilde w}_{\perp})} \delta(k'_{2-}) \delta(\tilde k_{2-})
\nonumber \\
&& \int d^2y_{\perp} \int_{-y_-^{(0)}}^{y_-^{(0)}} dy_- 
   \theta(y_- -w_-) \theta(y_- -{\tilde w}_-)
   \frac{(w_{\perp}-y_{\perp})^{\alpha}}{|w_{\perp}-y_{\perp}|^2}
   \frac{({\tilde w}_{\perp}-y_{\perp})^{\tilde\alpha}}
        {|{\tilde w}_{\perp}-y_{\perp}|^2} 
   \frac{\alpha_s}{2\pi N_c}\rho_N(y_{\perp},y_-) 
\delta^{a\tilde a}
\nonumber \\
&& \frac{2C_F N_c}{N_c^2-1} 
    \exp \left(-g^2\frac{\pi\rho_{\rm rel}N_c}{4(N_c^2-1)} 
      xG(x,|\tilde w_{\perp}-w_{\perp}|^{-2})
      ({\tilde w}_{\perp}-w_{\perp})^2(y_- +y_-^{(0)}) \right)
\label{eq16}
\end{eqnarray}

Next we perform the $y_-$ integration. 
To do so we make one more approximation. 
I assume $y_-^{(0)} \ll w_-,\tilde w_-$ and substitute 
the $\theta$-functions by $\theta (-w_-) \theta (-\tilde w_-)$.

\begin{eqnarray} 
\langle {\cal J}\rangle &=& V^{-2} \int d^2w_{\perp} 
  \int_{-\infty}^0 dw_- e^{-{\rm i}(k'_{2+}w_- -k'_{2\perp}\cdot w_{\perp})} 
  \int d^2\tilde w_{\perp}
\int_{-\infty}^0 d{\tilde w}_-
  e^{{\rm i}({\tilde k}_{2+}{\tilde w}_- -{\tilde k}_{2\perp}
     \cdot {\tilde w}_{\perp})} \delta(k'_{2-}) \delta(\tilde k_{2-})
\nonumber \\
&& \int d^2y_{\perp} \delta^{a \tilde a}
   \frac{(w_{\perp}-y_{\perp})^{\alpha}}{|w_{\perp}-y_{\perp}|^2}
   \frac{({\tilde w}_{\perp}-y_{\perp})^{\tilde\alpha}}
        {|{\tilde w}_{\perp}-y_{\perp}|^2}
   \frac{4(N_c^2-1)2y_-^{(0)}}{g^2\pi N_c\rho_{\rm rel}
   |\tilde w_{\perp}-w_{\perp}|^2 xG(x,|\tilde w_{\perp}-w_{\perp}|^{-2})}
\nonumber \\
&& \left[1-\exp \left(-g^2\frac{2y_-^{(0)}\pi\rho_{\rm rel}N_c}{4(N_c^2-1)}
   ({\tilde w}_{\perp}-w_{\perp})^2 xG(x,|\tilde w_{\perp}-w_{\perp}|^{-2}) 
   \right)\right] \frac{\alpha_s}{2\pi N_c}\rho_N(y_{\perp},y_-)
\label{eq19}
\end{eqnarray}
In the spirit of the discussion above we assume 
\begin{equation}
\frac{\rho_N(y_{\perp},y_-)}{2y_-^{(0)}\rho_{\rm rel}} = 1
\label{eq20}
\end{equation}
and use the relation (valid at leading logarithmic accuracy)
\begin{equation}
\frac{\partial xG(x,Q^2)}{\partial \ln Q^2} 
= \frac{\alpha_s(N_c^2-1)}{2\pi N_c} 
\qquad \longrightarrow \qquad  
xG(x,Q^2)= \frac{\alpha_s(N_c^2-1)}{2\pi N_c} \ln(Q^2/\mu^2)
\label{eq21}
\end{equation}
to obtain
\begin{eqnarray} 
\langle {\cal J}\rangle &=& V^{-2} \int d^2w_{\perp} 
  \int_{-\infty}^0 dw_- e^{-{\rm i}(k'_{2+}w_- -k'_{2\perp}\cdot w_{\perp})} 
  \int d^2{\tilde w}_{\perp} \int_{-\infty}^0 d{\tilde w}_-
  e^{{\rm i}({\tilde k}_{2+}{\tilde w}_- -{\tilde k}_{2\perp}
     \cdot {\tilde w}_{\perp})} 
\nonumber \\
&& \int d^2y_{\perp} 
   \frac{(w_{\perp}-y_{\perp})^{\alpha}}{|w_{\perp}-y_{\perp}|^2}
   \frac{({\tilde w}_{\perp}-y_{\perp})^{\tilde\alpha}}
        {|{\tilde w}_{\perp}-y_{\perp}|^2} \delta^{a \tilde a}
   \frac{-4}{g^2\pi N_c({\tilde w}_{\perp}-w_{\perp})^2
   \ln(\mu^2|\tilde w_{\perp}-w_{\perp}|^2)}
\nonumber \\
&& \left[1-\exp \left(-g^2\frac{2y_-^{(0)}\pi\rho_{\rm rel}N_c}{4(N_c^2-1)}
   ({\tilde w}_{\perp}-w_{\perp})^2 xG(x,|\tilde w_{\perp}-w_{\perp}|^{-2}) 
   \right)\right]\delta(k'_{2-}) \delta(\tilde k_{2-})
\label{eq22}
\end{eqnarray}
Next, using the two dimensional Green's function
\begin{equation}
\frac{(w_{\perp}-y_{\perp})^{\alpha}}{|w_{\perp}-y_{\perp}|^2} 
= -{\rm i}\int\frac{d^2q_{\perp}}{2\pi} e^{{\rm i} q_{\perp}
   \cdot(w_{\perp}-y_{\perp})}\frac{q_{\perp}^{\alpha}}{q_{\perp}^2}
\label{eq23}
\end{equation}
the $y_{\perp}$ integral can be performed:
\begin{eqnarray}
\int d^2y_{\perp} 
  \frac{(w_{\perp}-y_{\perp})^{\alpha}}{|w_{\perp}-y_{\perp}|^2}
  \frac{({\tilde w}_{\perp}-y_{\perp})^{\tilde\alpha}}
       {|{\tilde w}_{\perp}-y_{\perp}|^2}
&=& -\int d^2y_{\perp} 
    \int\frac{d^2q_{\perp}}{2\pi} \frac{d^2\tilde q_{\perp}}{2\pi}
    e^{{\rm i}[q_{\perp}\cdot(w_{\perp}-y_{\perp})
       +{\tilde q}_{\perp}\cdot({\tilde w}_{\perp}-y_{\perp})]}  
    \frac{q_{\perp}^{\alpha}}{q_{\perp}^2}
    \frac{\tilde q_{\perp}^{\alpha}}{\tilde q_{\perp}^2}
\nonumber \\
&=& \int d^2 q_{\perp}\,
    e^{{\rm i} q_{\perp}\cdot(w_{\perp}-\tilde w_{\perp})}
    \frac{q_{\perp}^{\alpha}q_{\perp}^{\tilde\alpha}}{(q_{\perp}^2)^2}
\nonumber \\
&=& \delta^{\alpha \tilde\alpha} 
    \int_0^\infty \frac{dq_{\perp}}{q_{\perp}} \int_0^{2\pi} d\phi\, 
    e^{{\rm i}q_{\perp}|w_{\perp}-{\tilde w}_{\perp}|\cos\phi} 
    \left( \begin{array}{c}
    \cos^2 \phi\\
    \sin^2 \phi
    \end{array} \right)
\nonumber \\
&=& 2\pi \delta^{\alpha \tilde\alpha} 
    \int_0^\infty \frac{dq_{\perp}}{q_{\perp}}\,, 
    \left( \begin{array}{c}
    J'_1(q_{\perp}|w_{\perp}-{\tilde w}_{\perp}|)\\
    J'_1(q_{\perp}|w_{\perp}-{\tilde w}_{\perp}|)
    +J_2(q_{\perp}|w_{\perp}-{\tilde w}_{\perp}|)\\
    \end{array} \right)
\label{eq24}
\end{eqnarray}
Up to the logarithmic divergence at $q_{\perp} \to 0$
this integral is independent of $|w_{\perp}-\tilde w_{\perp}|$ as one 
can see by substituting $q_{\perp}|w_{\perp}-\tilde w_{\perp}| \to z$. 
To regularize the IR divergence we introduce a lower integration 
boundary for $z$ in the form $\mu|w_{\perp}-{\tilde w}_{\perp}|$
and insert the finite $z \to 0$ limit of the Bessel functions:
$J'_0(0)=1, J_2(-)=0$, obtaining:
\begin{equation}
\delta^{\alpha \tilde\alpha} 
  \int_{\mu|w_{\perp}-{\tilde w}_{\perp}|}^{\infty} \frac{dz}{z} 
= -\frac{1}{2} \delta^{\alpha \tilde\alpha} 
  \ln(\mu^2|w_{\perp}-{\tilde w}_{\perp}|^2) .
\label{eq25}
\end{equation}
We thus finally end up with the result
\begin{eqnarray} 
\langle {\cal J}\rangle &=& V^{-2} \int d^2w_{\perp} 
  \int_{-\infty}^0 dw_- e^{-{\rm i}(k'_{2+}w_- -k'_{2\perp}\cdot w_{\perp})} 
  \int d^2{\tilde w}_{\perp}
\int_{-\infty}^0 d{\tilde w}_-
  e^{{\rm i}({\tilde k}_{2+}{\tilde w}_- -{\tilde k}_{2\perp}
    \cdot {\tilde w}_{\perp})} 
  \delta^{a\tilde a}\delta^{\alpha \tilde\alpha}
  \frac{2}{g^2N_c |\tilde w_{\perp}-w_{\perp}|^2}
\nonumber \\
&& \left[1-\exp \left(-g^2\frac{2y_-^{(0)}\pi\rho_{\rm rel}N_c}{4(N_c^2-1)}
   ({\tilde w}_{\perp}-w_{\perp})^2 xG(x,|\tilde w_{\perp}-w_{\perp}|^{-2}) 
   \right)\right]\delta(k'_{2-}) \delta(\tilde k_{2-})
\label{eq26}
\end{eqnarray}
I insert a suitable $\epsilon$-prescription to perform the integration
over $w_-$:
\begin{eqnarray} 
\langle {\cal J}\rangle 
&=& \frac{2\delta^{a \tilde a}\delta^{\alpha \tilde\alpha}}{g^2 N_c V^2}
    \int \frac{d^2w_{\perp} d^2{\tilde w}_{\perp}}
              {|{\tilde w}_{\perp}-w_{\perp}|^2}
    e^{{\rm i}(k_{2\perp}\cdot w_{\perp} -{\tilde k}_{2\perp}\cdot 
      {\tilde w}_{\perp})} 
    \int_{-\infty}^0 dw_- \int_{-\infty}^0 d{\tilde w}_-
    e^{-{\rm i}((k_{2+}+{\rm i}\epsilon)w_- 
      -({\tilde k}_{2+}-{\rm i}\epsilon){\tilde w}_-)}
\nonumber \\
&& \left[1-\exp \left(-g^2\frac{2y_-^{(0)}\pi\rho_{\rm rel}N_c}{4(N_c^2-1)}
   ({\tilde w}_{\perp}-w_{\perp})^2 xG(x,|\tilde w_{\perp}-w_{\perp}|^{-2}) 
   \right)\right]\delta(k'_{2-}) \delta(\tilde k_{2-})
\nonumber \\
&=& \frac{2\delta^{a \tilde a}\delta^{\alpha \tilde\alpha}}{g^2 N_c V^2}
    \int d^2w_{\perp} \frac{\rm i}{k'_{2+}+{\rm i}\epsilon}
    e^{{\rm i}k_{2\perp}\cdot w_{\perp}} 
    \int d^2\tilde w_{\perp}
    \frac{-\rm i}{{\tilde k}_{2+}-{\rm i}\epsilon} 
    e^{-{\rm i}{\tilde k}_{2\perp}\cdot {\tilde w}_{\perp}} 
    |\tilde w_{\perp}-w_{\perp}|^{-2}
\nonumber \\
&& 
\left[1-\exp \left(-g^2\frac{2y_-^{(0)}\pi\rho_{\rm rel}N_c}{4(N_c^2-1)}
   ({\tilde w}_{\perp}-w_{\perp})^2 xG(x,|\tilde w_{\perp}-w_{\perp}|^{-2}) 
   \right)\right]\delta(k'_{2-}) \delta(\tilde k_{2-})
\label{eq27}
\end{eqnarray}
Next we substitute $Z_{\perp}=w_{\perp}+\tilde w_{\perp}$ 
and $z_{\perp}=w_{\perp}-\tilde w_{\perp}$, giving
\begin{eqnarray} 
\langle {\cal J}\rangle 
&=& \frac{2\delta^{a \tilde a}\delta^{\alpha \tilde\alpha}}{g^2 N_c V^2}
    \frac{1}{(k'_{2+}+{\rm i}\epsilon)({\tilde k}_{2+}-{\rm i}\epsilon)}
    \int d^2Z_{\perp}
\int d^2z_{\perp}
    e^{{\rm i}[k'_{2\perp}\cdot(Z_{\perp}+z_{\perp})/2 
      -\tilde k_{2\perp}\cdot(Z_{\perp}-z_{\perp})/2]} 
\nonumber \\
&& \left[1-\exp \left(-g^2\frac{2y_-^{(0)}\pi\rho_{\rm rel}N_c z_{\perp}^2}
   {4(N_c^2-1)} xG(x,|z_{\perp}|^{-2}) \right)\right]
\delta(k'_{2-}) \delta(\tilde k_{2-})
\label{eq28}
\end{eqnarray}
Now we can insert the definition of the saturation scale from eq.~(17) 
of \cite{Yuri2}. In doing so we identify $r$ in that equation with  
$y_-^{(0)}$:
\begin{equation}
g^2\frac{2y_-^{(0)}\pi\rho_{\rm rel}N_c z_{\perp}^2}{4(N_c^2-1)}
   xG(x,|z_{\perp}|^{-2})
= \frac{Q_s^2z_{\perp}^2}{4} 
\label{eq29}
\end{equation}
This gives:
\begin{equation} 
\langle {\cal J}\rangle 
= \frac{2\delta^{a \tilde a}\delta^{\alpha \tilde\alpha} (2\pi)^2}{g^2 N_c V^2}
  \frac{\delta^2(k'_{2\perp}-{\tilde k}_{2\perp})}
       {(k'_{2+}+{\rm i}\epsilon)({\tilde k}_{2+}-{\rm i}\epsilon)}
  \int \frac{d^2z_{\perp}}{z_{\perp}^2}
  e^{{\rm i}k'_{2\perp}\cdot z_{\perp}} 
  \left[1-\exp \left(-\frac{Q_s^2 z_{\perp}^2}{4}\right)\right] \delta(k'_{2-}) \delta(\tilde k_{2-})   
\label{eq30}
\end{equation}
To do the $z_{\perp}$-integral we use the fact that the integral is well 
behaved at $z_{\perp}\to 0$ and we assume some $\epsilon$ 
prescription to make it convergent at $z_{\perp}\to \infty$.
\begin{equation}
K \equiv \int \frac{d^2z_{\perp}}{ z_{\perp}^2}
  e^{{\rm i}k'_{2\perp}\cdot z_{\perp}} 
  \left[1-\exp \left(-\frac{Q_s^2 z_{\perp}^2}{4}\right)\right]
= \frac{Q_s^2}{4} \int_0^1 du \int d^2z_{\perp}
  e^{{\rm i}k'_{2\perp}\cdot z_{\perp}} 
  \exp\left(-\frac{Q_s^2 z_{\perp}^2u}{4}\right)
\label{eq31}
\end{equation}
Substituting $z_{\perp}\to z_{\perp}+{\rm i}2k'_{2\perp}/(Q_s^2u)$, we get:
\begin{equation}
K = \frac{Q_s^2}{4} \int_0^1 du \int d^2z_{\perp}
    \exp\left(-\frac{Q_s^2 z_{\perp}^2u}{4}
              -\frac{{k'}_{2\perp}^2}{Q_s^2u}\right)
= \pi \int_0^1 \frac{du}{u} 
  \exp\left(-\frac{{k'}_{2\perp}^2}{Q_s^2u}\right)
\label{eq32}
\end{equation}
Next we substitute $u \to {k'}_{2\perp}^2/(Q_s^2t)$ to obtain:
\begin{equation}
K = \pi \int_{{k'}_{2\perp}^2/Q_s^2}^{\infty} 
\frac{dt}{t} e^{-t}
= \pi {\rm E}_1\left(\frac{{k'}_{2\perp}^2}{Q_s^2}\right) ,
\label{eq33}
\end{equation}
yielding finally 
\begin{equation}
\langle {\cal J}\rangle =
\frac{\delta^{a \tilde a}\delta^{\alpha \tilde\alpha}(2\pi)^3}{ g^2 N_c V^2}
  \frac{\delta^2(k'_{2\perp}-{\tilde k}_{2\perp})}
       {(k'_{2+}+{\rm i}\epsilon)({\tilde k}_{2+}-{\rm i}\epsilon)}
  {\rm E}_1\left( \frac{{k'}_{2\perp}^2}{Q_s^2} \right)\delta(k'_{2-}) \delta(\tilde k_{2-})
\label{eq34}
\end{equation}

thus we get:
\begin{eqnarray} 
{\cal W}_{1\hat 1,1'\tilde 1}&=& 
\int dk'_{2-} \int d \tilde k_{2-} \delta(k'_{2-}+k'_{1-}-k_{1-})
  \delta({\tilde k}_{2-}+{\tilde k}_{1-}-\hat k_{1-})  \delta(k'_{2-}) 
\delta(\tilde k_{2-}) 
\nonumber \\
&&\frac{(2\pi)^3}{V} f_{abc} f_{a{\tilde b} \hat c}
\int dk'_{2+}d^2k'_{2\perp} \theta(k'_{2+}) 
\frac{1}{\sqrt{2k_{1+}\hat k_{1+}}}
    \int d{\tilde k}_{2+}d^2{\tilde k}_{2\perp} \theta({\tilde k}_{2+}) 
    \delta(k'_{2+}+k'_{1+}-k_{1+})
\nonumber \\
&&  \delta^2(k'_{2\perp}+k'_{1\perp}-k_{1\perp})
\delta({\tilde k}_{2+}+{\tilde k}_{1+}-\hat k_{1+})
    \delta^2({\tilde k}_{2\perp}+{\tilde k}_{1\perp}-\hat k_{1\perp})
    \delta^2(k'_{2\perp}-{\tilde k}_{2\perp})
\nonumber \\
&& \left[ g_{\alpha\beta}\left(k'_2-k'_1\right)_{\gamma} 
  +g_{\beta\gamma}\left(k'_1+k_1\right)_{\alpha} 
  +g_{\gamma\alpha}\left(-k_1-k'_2\right)_{\beta} \right] 
\nonumber \\
&& \left.\left[ g^{~\alpha}_{\tilde\beta}
    \left(\tilde k_2-\tilde k_1\right)_{\tilde\gamma} 
  +g_{\tilde\beta\tilde\gamma}\left(\tilde k_1+\hat k_1\right)^{\alpha} 
  +g_{\tilde\gamma}^{~\alpha}\left(-\hat k_1-\tilde k_2\right)_{\tilde\beta} 
\right] \right|_{\alpha=1,2}
\nonumber \\
&& \frac{1}
       {(k'_{2+}+{\rm i}\epsilon)({\tilde k}_{2+}-{\rm i}\epsilon)}
  {\rm E}_1\left( \frac{{k'}_{2\perp}^2}{Q_s^2} \right)
\epsilon(k'_1)^{\beta} \epsilon^*(\tilde k_1)^{\tilde\beta}  
\Bigl( \sum_{\epsilon}  \epsilon(k_1)^{\gamma} \epsilon^*(\hat k_1)^{\tilde\gamma}  \Bigr)
\nonumber \\
&=& \frac{(2\pi)^3\delta^2(k'_{1\perp}-k_{1\perp}+\hat k_{1\perp}-{\tilde k}_{1\perp})
f_{abc} f_{a{\tilde b} \hat c}}
         {V} \delta(k'_{1-}-k_{1-})
  \delta({\tilde k}_{1-}-\hat k_{1-}) 
\nonumber \\
&&\int dk'_{2+}\theta(k'_{2+}) 
    \int d{\tilde k}_{2+}\theta({\tilde k}_{2+}) 
\frac{1}{\sqrt{2k_{1+}\hat k_{1+}}}
    \delta(k'_{2+}+k'_{1+}-k_{1+}) 
    \delta({\tilde k}_{2+}+{\tilde k}_{1+}-\hat k_{1+})
\nonumber \\
&& \left[ g_{\alpha\beta}\left(k'_2-k'_1\right)_{\gamma} 
  +g_{\beta\gamma}\left(k'_1+k_1\right)_{\alpha} 
  +g_{\gamma\alpha}\left(-k_1-k'_2\right)_{\beta} \right] 
\nonumber \\
&& \left.\left[ g^{~\alpha}_{\tilde\beta}
    \left(\tilde k_2-\tilde k_1\right)_{\tilde\gamma} 
  +g_{\tilde\beta\tilde\gamma}\left(\tilde k_1+\hat k_1\right)^{\alpha} 
  +g_{\tilde\gamma}^{~\alpha}\left(-\hat k_1-\tilde k_2\right)_{\tilde\beta} 
\right] \right|_{\alpha=1,2}
\nonumber \\
&& \frac{1}
       {(k'_{2+}+{\rm i}\epsilon)({\tilde k}_{2+}-{\rm i}\epsilon)}
  {\rm E}_1\left( \frac{{k'}_{2\perp}^2}{Q_s^2} \right)
\epsilon(k'_1)^{\beta} \epsilon^*(\tilde k_1)^{\tilde\beta}  
\Bigl( \sum_{\epsilon}  \epsilon(k_1)^{\gamma} \epsilon^*(\hat k_1)^{\tilde\gamma}  \Bigr)
\nonumber \\
&=& \frac{(2\pi)^3\delta^2(k'_{1\perp}-k_{1\perp}+\hat k_{1\perp}-{\tilde k}_{1\perp})
f_{abc} f_{a{\tilde b} \hat c}}
         {V
\sqrt{2k_{1+}\hat k_{1+}}
} \theta(\hat k_{1+}-\tilde k_{1+}) \theta(k_{1+}-k'_{1+}) 
  \delta(k'_{1-}-k_{1-}) \delta({\tilde k}_{1-}-\hat k_{1-})
\nonumber \\
&& \left[ g_{\alpha\beta}\left(k'_2-k'_1\right)_{\gamma} 
  +g_{\beta\gamma}\left(k'_1+k_1\right)_{\alpha} 
  +g_{\gamma\alpha}\left(-k_1-k'_2\right)_{\beta} \right] 
\nonumber \\
&& \left.\left[ g^{~\alpha}_{\tilde\beta}
    \left(\tilde k_2-\tilde k_1\right)_{\tilde\gamma} 
  +g_{\tilde\beta\tilde\gamma}\left(\tilde k_1+\hat k_1\right)^{\alpha} 
  +g_{\tilde\gamma}^{~\alpha}\left(-\hat k_1-\tilde k_2\right)_{\tilde\beta} 
\right] \right|_{\alpha=1,2}
\nonumber \\
&& \frac{1}
       {(k_{1+}-k'_{1+}+{\rm i}\epsilon)
(\hat k_{1+}-\tilde k_{1+}-{\rm i}\epsilon)}
  {\rm E}_1\left( \frac{{k'}_{2\perp}^2}{Q_s^2} \right)
\epsilon(k'_1)^{\beta} \epsilon^*(\tilde k_1)^{\tilde\beta}  
\Bigl( \sum_{\epsilon}  \epsilon(k_1)^{\gamma} \epsilon^*(\hat k_1)^{\tilde\gamma}  \Bigr)
\label{eq37}
\end{eqnarray}
with $k'_{2\perp}=k_{1\perp}-k'_{1\perp}$ 
and $\tilde k_{2\perp}=\hat k_{1\perp}-\tilde k_{1\perp}$.\\

\begin{figure}[tb]   
\resizebox{0.6\linewidth}{!}
          {\rotatebox{0}{\includegraphics{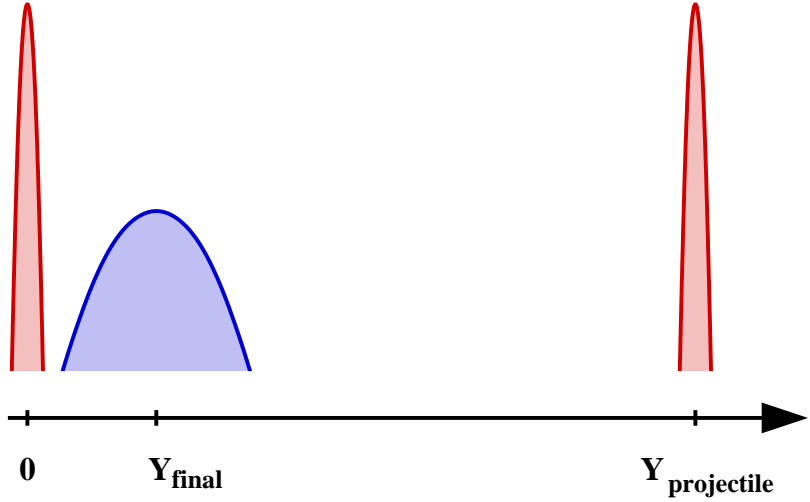}}}
\caption{The rapidity regions considered}
\label{fig2}
\end{figure}

The
WW-gluons are on-shell, i.e. $k'_{2-}=\tilde k_{2-}=0$ and thus
$k_{1-}=k'_{1-}$ and  $\hat k_{1-} =\tilde k_{1-}$. Therefore we can substitute in general
$k'_2=k_1-k'_1$  and $\tilde k_2=\hat k_1-\tilde k_1$.
 
\begin{eqnarray} 
{\cal W}_{1\hat 1,1'\tilde 1}
&=& \frac{(2\pi)^3\delta^2(k'_{1\perp}-k_{1\perp}+{\hat k}_{1\perp}
-{\tilde k}_{1\perp})
f_{abc} f_{a{\tilde b} \hat c}}
         {V
\sqrt{2k_{1+}\hat k_{1+}}
} 
~~\theta(\hat k_{1+}-\tilde k_{1+}) ~~\theta(k_{1+}-k'_{1+}) 
  \delta(k'_{1-}-k_{1-}) \delta({\tilde k}_{1-}-\hat k_{1-})
\nonumber \\
&& \left[ g_{\alpha\beta}\left(k_1-2k'_1\right)_{\gamma} 
  +g_{\beta\gamma}\left(k'_1+k_1\right)_{\alpha} 
  +g_{\gamma\alpha}\left(-2k_1+k'_1\right)_{\beta} \right] 
\nonumber \\
&& \left.\left[ g_{~\alpha}^{\tilde\beta}
    \left(\hat k_1-2\tilde k_1\right)_{\tilde\gamma} 
  +g_{\tilde\beta\tilde\gamma}\left(\tilde k_1+\hat k_1\right)^{\alpha} 
  +g_{\tilde\gamma}^{~\alpha}\left(-2 \hat k_1+\tilde k_1\right)_{\tilde\beta} 
\right] \right|_{\alpha=1,2}
\nonumber \\
&& \frac{1}
       {(k_{1+}-k'_{1+}+{\rm i}\epsilon)
(\hat k_{1+}-\tilde k_{1+}-{\rm i}\epsilon)}
  {\rm E}_1\left( \frac{{k'}_{2\perp}^2}{Q_s^2} \right)
\epsilon(k'_1)^{\beta} \epsilon^*(\tilde k_1)^{\tilde\beta}  
\Bigl( \sum_{\epsilon}  \epsilon(k_1)^{\gamma} \epsilon^*(\hat k_1)^{\tilde\gamma}  \Bigr)
\label{eq41}
\end{eqnarray}
At this point we have to make an assumption about how far in rapidity 
a gluon of nucleus 1 is scattered. Let us require an average rapidity 
$Y>1$ of the scattered gluons (see Fig.~\ref{fig2}). 
Decoherence will be effective, when $Y$ is large enough that the final 
phase space is originally empty or only sparsely occupied, which is the 
case for $Y>1$. This condition also implies $\hat k_{1+},k_{1+}\gg k'_{1+}, 
\tilde k_{1+}$. Decoherence may also occur inside the target region 
($Y\leq 1$), but we are not concerned with this question here.

With these assumptions it is now possible to greatly simplify 
Eq. (\ref{eq41}). In the square brackets we keep only $k_{1,0}$, $\hat k_{1,0}$ 
and $k_{1,3}$, $\hat k_{1,3}$. We neglect $k_{1,\alpha}$, because $\alpha$ denotes
a transverse direction. We can also drop $k_{1,\gamma}$ and 
$k_{1,\tilde\gamma}$ because of the projector at the end of 
Eq. (\ref{eq41}). Thus we are left with
\begin{eqnarray}  
2g_{\gamma\alpha}k_{1,\beta} 
g_{\tilde\gamma}^{~\alpha}2 \hat k_{1,\tilde\beta}  
\epsilon(k'_1)^{\beta} \epsilon^*(\tilde k_1)^{\tilde\beta}  
\left.
\left(g^{\gamma\tilde\gamma}-\frac{k_1^{\gamma} \hat k_1^{\tilde\gamma}}
    {k_1\cdot \hat k_1} \right) \right|_{\alpha=1,2} 
&=&8~ k_1\cdot \epsilon(k'_1) 
~\hat k_1\cdot \epsilon^*(\tilde k_1)
\left(1+\frac{k_{1,\perp}\cdot \hat k_{1,\perp}}
    {2k_1\cdot \hat k_1} \right)
\nonumber \\
&=&8~ k_{1+} \hat k_{1+}
\epsilon(k'_1)_- 
~\epsilon^*(\tilde k_1)_-
\left(1+\frac{k_{1,\perp}\cdot \hat k_{1,\perp}}
    {2k_1\cdot \hat k_1} \right)
\label{eq42}
\end{eqnarray}  
Making use of:
\begin{equation}
\sum_{\epsilon}  \epsilon(k_1)^{\gamma} \epsilon^*(\hat k_1)^{\tilde\gamma}  
=
\left(g^{\gamma\tilde\gamma}-\frac{k_1^{\gamma} \hat k_1^{\tilde\gamma}}
    {k_1\cdot \hat k_1} \right) 
\end{equation}

The next task is to evaluate the correlator
\begin{equation}  
\langle \epsilon(k'_1)_- 
~\epsilon^*(\tilde k_1)_-
\rangle
\label{eq43}
\end{equation}  
Next we have to evaluate the convolution with the density matrix
of the product of polarization vectors and the color factor:
\begin{equation}
\langle f_{abc} f_{a{\tilde b} \hat c}
\epsilon(k'_1)_- 
~\epsilon^*(\tilde k_1)_-
\rangle_{\rm polarization,color}
\sim \sum_{\epsilon} \sum_{b, \tilde b} \epsilon(k'_1)_- 
~\epsilon^*(\tilde k_1)_-
f_{abc} f_{a{\tilde b} \hat c} 
D\Bigl( b,\tilde b, \epsilon(k'_1),\epsilon(\tilde k'_1);
k'_1,\tilde k_1 \Bigr)
\end{equation}  
To do so we Fourier transform into coordinate space and simply assume 
that there (in the rest frame of nucleus 1)  the correlation of the 
polarization vectors and colors is given by a simple Gaussian with correlation 
length $\lambda$, i.e. we assume that $D$ is a
Gaussian in coordinate space.  
\begin{eqnarray}  
\langle f_{abc} f_{a{\tilde b} \hat c}
\epsilon(k'_1)_- 
~\epsilon^*(\tilde k_1)_-
~\rangle
&=& 
C\delta_{c\hat c}  \int ~d^2 y'_{\perp} ~d^2 \tilde y_{\perp} ~dy'_- ~d\tilde y_- 
~dy'_+ ~d\tilde y_+ 
\nonumber \\
&&
\exp\Big\{{\rm i} (k'_{1,\perp}\cdot y'_{\perp}- 
\tilde k_{1,\perp}\cdot \tilde y_{\perp}
-k'_{1,+} y'_- + \tilde k_+\tilde y_-)
-k'_{1,-} y'_+ + \tilde k_-\tilde y_+)\Big\} ~
\nonumber \\
&&
\exp\Big\{-[(y'_{\perp}-\tilde y_{\perp})^2
+(y'_-- \tilde y_-)^2+(y'_+- \tilde y_+)^2]/\lambda^2\Big\} 
\nonumber \\
&=&
\frac{C\delta_{c\hat c}}{16} \int ~d^4 Y^{(+)} ~d^4 Y^{(-)}- 
\exp\Big\{-\frac{\rm i}{2} [(k'_{1}- \tilde k_{1})
\cdot Y^{(+)}+(k'_{1}+ \tilde k_{1})\cdot Y^{(-)}] \Big\} ~
\nonumber \\
&&
\exp\Big\{[(Y^{(-)}_+)^2+(Y^{(-)}_-)^2+(Y^{(-)}_{\perp})^2]/\lambda^2\Big\} 
\nonumber
\\
&=&C\delta_{c\hat c} 
(\sqrt{\pi}\lambda)^4
\delta^2(k'_{1,\perp}-\tilde k_{1,\perp})
\delta(k'_{1-}-\tilde k_{1-}) \delta(k'_{1+}-\tilde k_{1+})
(2\pi)^4
\nonumber \\
&&
\exp\Big\{-\lambda^2[(k'_{1,\perp}+\tilde k_{1,\perp})^2+
(k'_{1-}+\tilde k_{1-})^2 +(k'_{1+}+\tilde k_{1+})^2]/4\Big\} 
\nonumber \\
&=&C\delta_{c\hat c} 
(\sqrt{\pi}\lambda)^4
\delta^2(k'_{1,\perp}-\tilde k_{1,\perp})
\delta(k'_{1-}-\tilde k_{1-}) \delta(k'_{1+}-\tilde k_{1+})(2\pi)^4
\nonumber \\
&&
\exp\Big\{-\lambda^2[(k'_{1,\perp})^2+
(k'_{1-})^2+(k'_{1+})^2]\Big\} 
\label{eq45}
\end{eqnarray}
with $Y^{(+)}=y'+\tilde y$ and $Y^{(-)}=y'-\tilde y$. $C$ is a 
constant to be determine from a suitable normalization condition.\\

Now we are at the point where the normalization of $d$, respectively $D$,  
has to be discussed. To do so we start from the energy associated with the
quadratic part of the Yang-Mills Lagrangian,
\begin{equation}
{\cal L}= -\frac{1}{2} (\partial_{\mu}A_{\nu}^a-\partial_{\nu}A_{\mu}^a)
\partial^{\mu}A_{a\nu} ,
\label{D1}
\end{equation}
which should be equivalent to some integral of the form
\begin{equation}
E= \int \frac {d^4p}{(2\pi)^2}~~ ... ~~ d(p,p)
\label{D2}
\end{equation}
with a one-particle, not yet normalized, density matrix $d(p,p)$.
>From Eq. (\ref{D1}) we get 
\begin{equation}
E= \frac {1}{2}\int d^3x ~\Bigl( (\partial^0A^{bj}(x))^2-(\partial^jA^{b0}
(x))^2
+(\partial^iA^{bj}(x))(\partial^iA^{bj}(x))
-(\partial^iA^{bj}(x))(\partial^jA^{bi}(x)) \Bigr)
\label{D3}
\end{equation}
Substituting 
\begin{equation}
A^{b\mu}(x)A^{b\nu}(x)~~\to~~
\int \frac {d^4p}{(2\pi)^4}~\int \frac {d^4 \tilde p}{(2\pi)^4}
e^{{\rm i}(p-\tilde p)\cdot x} ~d^{\mu\nu}(b,b;p;\tilde p)
\end{equation}
we get 
\begin{equation}
E= \frac {1}{2}
\int \frac {d^4p~d\tilde p^0}{(2\pi)^5}
~\Bigl( (p^0 \tilde p^0 +(\vec p)^2)
d^{jj}(b,b;p^0,\vec p;\tilde p^0,\vec p)
-(\vec p)^2 d^{00}(b,b;p^0,\vec p;\tilde p^0,\vec p)
-p^i{\tilde p}^j d^{ji}(b,b;p^0,\vec p;\tilde p^0,\vec p)
\Bigr)
\label{D4}
\end{equation}
For simplicity we choose the gauge $p_\mu A^\mu =0$, allowing the last term 
to be re-expressed in the form $p^0 \tilde p^0 d^{00}$. 
Our model assumptions Eq. (\ref{eq45}) imply that we only need the diagonal 
matrix elements of $D$. To fix it we require that  Eq. (\ref{D4})
reproduces $\bar p^0$ for a plane wave gluon with 4 momentum $\bar p^\mu$,
i.e. we impose:
\begin{equation}
\bar p^0= \frac {1}{2}
\int \frac {d^4p~d\tilde p^0}{(2\pi)^5}
~\Bigl( 
(p^0 \tilde p^0 +(\vec p)^2)
d^{jj}_{\bar p}(b,b;p^0,\vec p,\tilde p^0,\vec p)
-(p^0 \tilde p^0 +(\vec p)^2)d^{00}_{\bar p}(b,b;p^0,\vec p,\tilde p^0,\vec p)
\Bigr)
\label{D5}
\end{equation}
for a density matrix of the form 
\begin{equation}
d^{\mu\nu}_{\bar p}(b,b;p^0,\vec p,\tilde p^0,\vec p)=
\delta_{\mu k} \delta_{\nu k} (2\pi)^4\delta^4(p-\bar p)\, f(p^0, \tilde p^0,\vec p)
\label{D6}
\end{equation}
The solution for $f$ is obviously 
\begin{equation}
f(p^0, \tilde p^0,\vec p)=\delta(p^0-\tilde p^0)\frac{p^04\pi}{(p^0)^2
+(\vec p)^2}
\end{equation}
This motivates our assumption that 
the diagonal matrix elements of our $d$ are related to those of the 
properly normalized density matrix $D$ by the factor (setting $\delta(0)=T$)
\begin{equation}
d^{\mu\nu}(b,b;p^0,\vec p,p^0,\vec p)=
T~\frac{p^04\pi}{(p^0)^2
+(\vec p)^2}~~D^{\mu\nu}(b,b;p^0,\vec p,p^0,\vec p)~VT ,
\end{equation}
where $D$ is according to Eq. (\ref{eq45}) in our model given by
\begin{equation}
D^{\mu}_{~~\nu}(b,b;p^0,\vec p,p^0,\vec p) =
\frac{16}{VT}\pi^2\lambda^4
\exp\Big\{-\lambda^2[(p_{\perp})^2+(p_{-})^2+(p_{+})^2]\Big\} 
\frac{1}{2}\delta_{\mu\nu}|_{\mu,\nu  \neq \perp} .
\end{equation}
Here the superscript ``$\perp$'' indicates that the Kronecker symbol
contributes only for the transverse directions $\mu,\nu=1,2$, and
the factor $\frac{1}{2}$ encodes the unpolarized nature of the gluons
in the target nucleus.
The normalization of the trace of the density matrix demands that
\begin{equation}
VT~\int \frac {d^4p}{(2\pi)^4}
D^{\mu}_{~~\mu}(b,b;p^0,\vec p,p^0,\vec p)=1
\label{eq54}
\end{equation}
for a one-gluon state.

For a many-gluon state this should be normalized to the total number of 
gluons which is an ill-defined quantity. We therefore choose to 
substitute for this general case $E=\bar p^0$ by the total energy of 
the gluons $E_G(Q_s^2)$ in nucleus 1, at the transverse scale with which this 
nucleus is resolved, which is the saturation scale.

Combining everything the density matrix in the boosted, previously 
unpopulated region of phase space is 
\begin{eqnarray}
\frac{1}{2}\delta_{\mu\nu}|_{\mu,\nu  \neq \perp} .
D_{1\hat 1}(c,\hat c;k_1^0,\vec k_1,\hat k_1^0,\vec{\hat  k}_1)
&=&
\int \frac{d^4{\tilde k}_1}{(2\pi)^4}~
\int \frac{d^4k_1'}{(2\pi)^4}~(2\pi)^4\delta^4(k_1'-{\tilde k}_1)\frac{1}{VT} 
\nonumber \\
&& \qquad\qquad
{\cal W}_{1\hat 1,\tilde 1\tilde 1}
~\frac{4\pi E_G(Q_s^2)T}{({\tilde k}_1^0)^2+(\vec{\tilde k}_1)^2}
~~D_{\mu\nu}(b,b;k'_1,{\tilde k}_1)~VT
\end{eqnarray}
with
\begin{equation}
f_{abc} f_{ab \hat c}\,
\epsilon(k'_1)_- \, \epsilon^*(k'_1)_- \,
f_{abc} f_{ab \hat c}
~~\to~~
N_c \delta_{c\hat c}\frac{1}{3} ~~~.
\end{equation}

The factor $\frac{1}{3}$ is motivated by the assumption that in a nucleus 
at rest (the target nucleus 1) all gluons are so highly virtual that the 
transverse and longitudinal polarization components contribute equally. 
The gauge degrees of freedom, of course, do not contribute.

Combining all our results we get for the one particle density matrix in the 
final state phase space region (substituting $\tilde k_1$ by $p$ for 
notational simplicity):
\begin{eqnarray}
D_{1,\hat 1}&=&
\int \frac{d^4 p}{(2\pi)^4}~
\frac{4\pi E_G(Q_s^2) 16 \pi^2 \lambda^4 N_c\delta_{c\hat c}}
{3V^2((p^0)^2+(\vec p)^2)}
\exp\Big\{-\lambda^2[(p_{\perp}^2+p_-^2+p_+^2]\Big\}
\nonumber \\ 
&&\frac{(2\pi)^3\delta^2(k_{1\perp}-{\hat k}_{1\perp})}
{\sqrt{2k_{1+}\hat k_{1+}}} 
~~\theta(\hat k_{1+}-p_+) ~~\theta(k_{1+}-p_+) 
  \delta(p_--k_{1-}) \delta(p_--\hat k_{1-})
\nonumber \\
&&
8~ k_{1+} \hat k_{1+}
\left(1+\frac{k_{1,\perp}\cdot \hat k_{1,\perp}}
    {2k_1\cdot \hat k_1} \right)
\frac{1}{(k_{1+}-p_++{\rm i}\epsilon)
(\hat k_{1+}-p_+-{\rm i}\epsilon)}
  {\rm E}_1\left( \frac{(k_{1\perp}-p_{\perp})^2}{Q_s^2} \right)
\label{eq57}
\end{eqnarray}
In order to evaluate this expression, we note that we are interested
in the scattering of gluons into states with $k_{1+},{\hat k}_{1+}
\gg p_+ \sim {\cal O}(\lambda^{-1})$. We can then neglect the $p_+$
dependence of the two denominators, drop the step functions, and obtain:
\begin{eqnarray}
&& 16\pi^2\int \frac{d^4 p}{(2\pi)^4}~
\frac{e^{-\lambda^2[p_{\perp}^2+p_-^2+p_+^2]}}{p_{\perp}^2+p_-^2+p_+^2}~
\delta(p_--k_{1-}) \delta(p_--\hat k_{1-})~
{\rm E}_1\left( \frac{(k_{1\perp}-p_{\perp})^2}{Q_s^2} \right)
\nonumber \\
&& \qquad =
\frac{1}{\pi^2} \delta(k_{1-}-\hat k_{1-}) 
\int_{\lambda^2}^{\infty} d\xi^2 \int_{-\infty}^{\infty} dp_+ \int_{-\infty}^{\infty} d^2p_{\perp}
e^{-\xi^2(p_+^2+p_{\perp}^2+k_{1-}^2)}
{\rm E}_1\left( \frac{(k_{1\perp}-p_{\perp})^2}{Q_s^2} \right)
\nonumber \\
&& \qquad =
\frac{2}{\pi^{3/2}} \delta(k_{1-}-\hat k_{1-}) 
\int_{\lambda}^{\infty} d\xi \int_0^1 \frac{du}{u} \int d^2p_{\perp}
e^{-\xi^2 (p_{\perp}^2+k_{1-}^2)}~e^{-(k_{1\perp}-p_{\perp})^2/(Q_s^2u)}
\nonumber \\
&& \qquad =
\frac{2}{\sqrt{\pi}} \delta(k_{1-}-\hat k_{1-}) 
\int_{\lambda}^{\infty} d\xi~ e^{-\xi^2 k_{1-}^2}
\int_0^1 du~ \frac{Q_s^2}{uQ_s^2\xi^2+1}~
\exp\left( -\frac{\xi^2 k_{1\perp}^2}{uQ_s^2\xi^2+1}\right) 
\end{eqnarray}
We now have traded 3 integrals (over $p_+$ and $p_{\perp}$) for two
integrals (over $\xi$ and $u$). This may not seem like much progress,
but it turns out that the integral over $u$ can be done after the
substitution $s=\xi^2Q_s^2/(u\xi^2Q_s^2+1)$:
\begin{eqnarray}
\int_0^1 du~ \frac{1}{uQ_s^2\xi^2+1}~
\exp\left( -\frac{\xi^2 k_{1\perp}^2}{uQ_s^2\xi^2+1}\right) 
& =
&\int_{\frac{Q_s^2\xi^2}{Q_s^2\xi^2+1}}^{\xi^2Q_s^2} \frac{ds}{s^2}
\frac{s}{\xi^2Q_s^2}
e^{-sk_{1\perp}^2/Q_s^2}
\nonumber \\
& = &
\frac{1}{Q_s^2\xi^2}
\left[  
E_1\left(\frac{k_{1\perp}^2\xi^2}{Q_s^2\xi^2+1}\right)
  - E_1(k_{1\perp}^2\xi^2) 
\right]  .
\end{eqnarray}

We finally substitute $\xi \to \lambda\xi$ in the remaining 
integration and obtain for the expression (\ref{eq57})
\begin{eqnarray}
D_{1,\hat 1} &=& 
\frac{(4\pi)^4 \lambda^3 E_G N_c \delta_{c\hat c}}{3 V^2}
\frac{\delta^2(k_{1\perp}-{\hat k}_{1\perp})\delta(k_{1-}-{\hat k}_{1-})}
     {\sqrt{2k_{1+}{\hat k}_{1+}}}
\left(1+\frac{k_{1,\perp}\cdot \hat k_{1,\perp}}{2k_1\cdot \hat k_1}\right)
\nonumber \\
&& \times
\frac{2}{\sqrt{\pi}} 
\int_{1}^{\infty} \frac{d\xi}{\xi^2}~ e^{-\lambda^2 k_{1-}^2\xi^2}
\left[ E_1\left(\frac{\lambda^2k_{1\perp}^2\xi^2}{\lambda^2Q_s^2\xi^2+1}\right)
     - E_1(\lambda^2k_{1\perp}^2\xi^2) \right] .
\label{eq60}
\end{eqnarray}
This expression describes the density matrix of the liberated gluons,
which are scattered out of the target nucleus 1 by the quasi-real 
gluons of the fast moving projectile nucleus 2. We note that the
density matrix is diagonal in the momentum components $k_{1-}$ and
$k_{1\perp}$, but not in the component $k_{1+}$. The physical reason
for this asymmetric behavior is that the projectile nucleus is moving
very fast in the $x_{-}$ direction. This implies that the distribution
of its gluons in $k_{+}$ is very broad and leads to interference of
excitation amplitudes of gluons from the target nucleus into final 
states with different values of $k_{1+}$.

\section{The decoherence time}

We now calculate the ratio (\ref{eqTrace}) for the density matrix 
from Eq. (\ref{eq60}).
we define
\begin{equation}
F(k_{1-},k_{1\perp}) 
= \int_{1}^{\infty} \frac{d\xi}{\xi^2}~ e^{-\lambda^2 k_{1-}^2\xi^2}
\left[ E_1\left(\frac{\lambda^2k_{1\perp}^2\xi^2}
                     {\lambda^2Q_s^2\xi^2+1}\right)
     - E_1(\lambda^2k_{1\perp}^2\xi^2) \right]
\end{equation}

\begin{figure}[tb]   
\resizebox{0.6\linewidth}{!}
          {\rotatebox{0}{\includegraphics{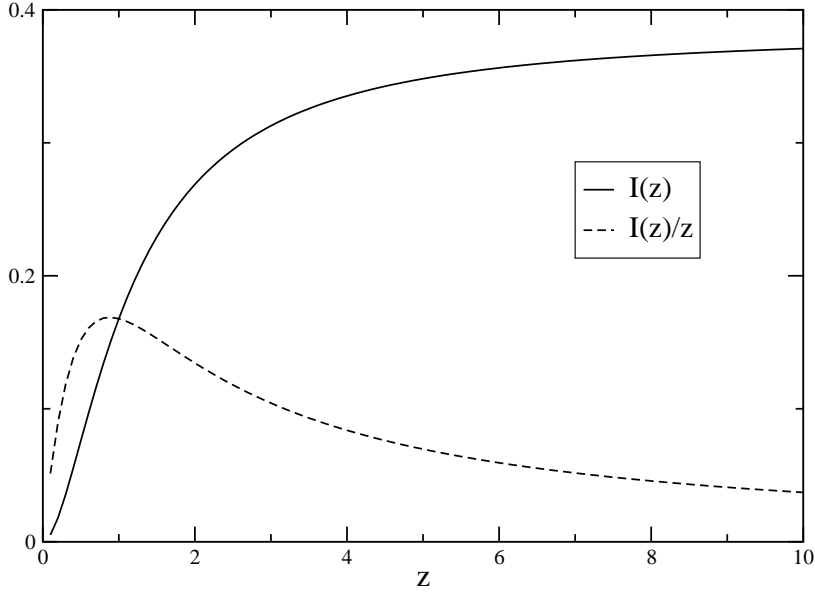}}}
\caption{The function $I(z)$ and the ratio $I(z)/z$}
\label{fig3}
\end{figure}

Because any constant factors will drop out of the ratio (\ref{eqTrace}),
it is sufficient to consider the $k$-dependent part of $D_{1,\hat 1}$,
which we call the reduced density matrix:
\begin{equation}
{\cal D}_{1,\hat 1} =
\frac{F(k_{1-},k_{1\perp})}{\sqrt{k_{1+}{\hat k}_{1+}}}
\delta(k_{1-}-{\hat k}_{1-})\delta^2(k_{1\perp}-{\hat k}_{1\perp}) ,
\end{equation}
where we have neglected the factor $(1+\ldots)$ deriving from the
polarization sum, which is of order unity. We obtain:
\begin{eqnarray}
{\rm Tr} {\cal D}
& = & \frac{VT}{(2\pi)^4}\delta(0_{-})\delta^2(0_{\perp})
\int \frac{dk_{1+}}{k_{1+}} \int_0^{\infty}dk_{1-} \int _{-\infty}^{\infty}d^2k_{1\perp} 
F(k_{1-},k_{1\perp})
\nonumber \\
& = & \frac{VT}{(2\pi)^4}\delta(0_{-})\delta^2(0_{\perp})
\int \frac{dk_{1+}}{k_{1+}} \frac{\pi^{3/2}}{4\lambda}Q_s^2
\end{eqnarray}
The details of the integration can be found in appendix A.
For ${\rm Tr}{\cal D}^2$ we obtain
\begin{eqnarray}
{\rm Tr}{\cal D}^2 &=&
\left( \frac{VT}{(2\pi)^4} \right)^2 
\int d^4k \int d^4\hat k \frac{1}{k_{1+}\hat k_{1+}}
\nonumber \\
&& F(k_{1-},k_{\perp}) \delta(k_{1-}-\hat k_{1-})  
   \delta^2(k_{1\perp}-\hat k_{1\perp})   
   F(k_{1-},k_{\perp}) \delta(k_{1-}-\hat k_{1-})  
   \delta^2(k_{1\perp}-\hat k_{1\perp})   
\nonumber \\
& = & \left( \frac{VT}{(2\pi)^4} \right)^2 
      \int \frac{dk_{1+}}{k_{1+}} \int \frac{d \hat k_{1+}}{\hat k_{1+}} \,
      \delta(0_{-})\delta^2(0_{\perp})  
      \int_0^{\infty}dk_{1-} \int _{-\infty}^{\infty}d^2k_{1\perp} 
      \left( F(k_{1-},k_{\perp})\right)^2
\end{eqnarray}
Which can be simplyfied using the identity (see Appendix A)
\begin{equation}
\int _{-\infty}^{\infty}dk_{1\perp}^2E_1(a k_{1\perp}^2)E_1(b k_{1\perp}^2)
= \frac{1}{b}\ln \frac{a+b}{a}+ \frac{1}{a}\ln \frac{a+b}{b}
\end{equation}
which leads to 
\begin{eqnarray}
{\rm Tr}{\cal D}^2 &=&
\left( \frac{VT}{(2\pi)^4} \right)^2 
\int \frac{dk_{1+}}{k_{1+}} \int  \frac{d \hat k_{1+}}{\hat k_{1+}} \
\delta(0_{-})\delta^2(0_{\perp}) 
\frac{\pi\sqrt{\pi}}{2\lambda}
Q_s^2
\int_1^{\infty}\frac{d\xi}{\xi^2} \int_1^{\infty}\frac{d\chi}{\chi^2} 
\frac{1}{\sqrt{\xi^2+\chi^2}} 
\nonumber \\
&& \qquad
\sum_{i,j=1}^2~ (-1)^{i+j}\left[ 
\frac{1}{b_j}\ln \frac{a_i+b_j}{a_i}+ \frac{1}{a_i}\ln \frac{a_i+b_j}{b_j}
\right]
\end{eqnarray}
with 
\begin{eqnarray}
a_1=\frac{\lambda^2Q_s^2\xi^2}{\lambda^2Q_s^2\xi^2+1} &~ ~ ~,~ ~ ~&
a_2=\lambda^2Q_s^2\xi^2
\nonumber  \\
b_1=\frac{\lambda^2Q_s^2\chi^2}{\lambda^2Q_s^2\chi^2+1} &~ ~ ~,~ ~ ~&
b_2=\lambda^2Q_s^2\chi^2
\end{eqnarray}
We now define the integral 
\begin{equation}
I(\lambda Q_s)= \int_1^{\infty}\frac{d\xi}{\xi^2} 
 \int_1^{\infty}\frac{d\chi}{\chi^2} \frac{1}{\sqrt{\xi^2+\chi^2}} 
 \sum_{i,j=1}^2~ (-1)^{i+j}\left[ 
 \frac{1}{b_j}\ln \frac{a_i+b_j}{a_i}+ \frac{1}{a_i}\ln \frac{a_i+b_j}{b_j}
 \right] ,
\end{equation}
which approaches zero for $\lambda Q_s \rightarrow 0$ 
and approaches for asymptotically  large argument
the limit $I(\infty) = \frac{4}{3}(\sqrt{2}-1)\ln 2
\approx 0.3828$. One can see the latter as follows: 
$\lambda Q_s$ is much larger than unity and also $\xi, \chi \geq 1$. 
Therefore $a_2\gg a_1 \sim 1$ and $b_2\gg b_1\sim 1 $ and the contribution 
with $i=j=1$ dominates. In this limit, the square bracket assumes a value 
of about $2\ln 2$. Substituting this value into the integral (68), we obtain
the asymptotic limit mentioned above. 
The exact form of $I(\lambda Q_s)$ is given in Fig. \ref{fig3}.

The degree of decoherence is given by the ratio
\begin{equation}
\frac{{\rm Tr}{\cal D}^2(total)}{\left( {\rm Tr}{\cal D}(total) \right)^2}
= 
\frac{{\rm Tr} [{\cal D}(target) + {\cal D}]^2 }
{\left( {\rm Tr}[{\cal D}(target) + {\cal D}]\right)^2}
\label{ratio1}
\end{equation}
where ${\cal D}(target)$ is the density matrix of that part of the phase space 
which contained the original target.   
We are interested in the situation that most gluons of nucleus 1
have already undergone a hard scattering, i.e. $T$ must be large enough, typically 
of order $\ge 0.1$ fm/c. Then we can disregard
${\cal D}(target)$.

Thus we finally obtain for the ratio which characterizes the degree of decoherence
after the hard gluon scattering  
\begin{equation}
\frac{{\rm Tr}{\cal D}^2}{\left( {\rm Tr}{\cal D} \right)^2}
= \frac{\pi\sqrt{\pi}}{2\lambda} 
  \frac{16\lambda^2I(\lambda Q_s)Q_s^2}
       {\pi^3 Q_s^4 \delta^2(0_{\perp})\delta(0_-)}
= \frac{8I(\lambda Q_s)\lambda}{\pi\sqrt{\pi}Q_s^2} \cdot
\frac{\pi}{\lambda^2} \cdot \frac{1}{T}
\label{ratio}
\end{equation}
With the ``observation time'' $\delta(0_-)=T$ and 
$\delta^2(0_{\perp})=\lambda^2/\pi$, which is the correct 
normalization for eq.(\ref{eq54}), because
\begin{equation}
\delta^4(0)=\frac{VT}{(2\pi)^4} D_{\mu}^{\mu}(0)= \frac{\lambda^4}{\pi^2}.
\end{equation}
We now define the decoherence time as the value of $T$ for which the 
ratio (\ref{ratio}) becomes equal to $1/e$. This gives
\begin{equation}
{\tau}_{\rm deco} = \frac{8eI(\lambda Q_s)}{\sqrt{\pi}Q_s\lambda} \cdot 
\frac{1}{Q_s}
\end{equation}
For realistic values ($\lambda=0.3$~fm, $Q_s=1$~GeV) the first factor 
is numerically of order unity  (more precisely it 
is 1.865) 
for these values) and we can conclude 
\begin{equation}
{\tau}_{\rm deco} \sim  \frac{1}{Q_s}
\label{endres}
\end{equation}
This is our main result. The decoherence time is of the order of 
0.3-0.4 fm/c, and thus large enough to neglect 
${\cal D}(target)$ in Eq. (\ref{ratio1}),
and drops  with increasing saturation scale 
$Q_s$, i.e. with increasing 
collision energy. However, $Q_s$ increases so slowly with $s$
that the latter effect is rather marginal.
We use $\lambda Q_s$=1.5 and think that the uncertainty in this 
value is not large. Nevertheless, we want to point out that 
even if one varied this product by a factor of two, the ratio
$I(\lambda Q_s)/\lambda Q_s$ would change only marginaly
and (\ref{endres}) stayed valid.
(For $\lambda Q_s$ much smaller than one our description stops 
making sense.)

\section{Conclusions}

We have calculated the characteristic decoherence time in high energy 
heavy ion collisions due to gluon scattering. We find that this time 
is substantially shorter than 1 fm/$c$. This result furnishes the
remaining logical link in our argument that decoherence alone can 
explain a substantial part of the entropy production during the earliest 
phase of a heavy ion collison. We note that the decohered partonic 
state of the colliding nuclei is not yet thermally equilibrated.
Additional interactions among the decohered quanta, such as those
invoked in the bottom-up scenario of equilibration \cite{Baier} or
modified versions of it \cite{Bodeker,Mueller} are required to 
achieve full equilibration.

\section*{Acknowledgments:}

One of us (BM) gratefully acknowledges support from the Alexander von 
Humboldt Foundation in the form of a Senior Scientist Award. This work 
was also supported in part by grants from the Office of Science of the 
U.~S.~Department of Energy and from the BMBF.

\section{Appendix A}

We here evaluate the integral appearing in eq.~(60):
\begin{eqnarray}
I'&=& 
 \int_0^{\infty}dk_{1-} \int _{-\infty}^{\infty}d^2k_{1\perp} 
\int_{1}^{\infty} \frac{d\xi}{\xi^2}~ e^{-\lambda^2 k_{1-}^2\xi^2}
\left[ E_1\left(\frac{\lambda^2k_{1\perp}^2\xi^2}
                     {\lambda^2Q_s^2\xi^2+1}\right)
     - E_1(\lambda^2k_{1\perp}^2\xi^2) \right]
\nonumber \\
&=& 
\frac{\sqrt{\pi}}{2\lambda} 
\int _{-\infty}^{\infty}d^2k_{1\perp}
\int_{1}^{\infty} \frac{d\xi}{\xi^3}
\left[ E_1\left(\frac{\lambda^2k_{1\perp}^2\xi^2}
                     {\lambda^2Q_s^2\xi^2+1}\right)
     - E_1(\lambda^2k_{1\perp}^2\xi^2) \right]
\nonumber \\
&=& 
\frac{\sqrt{\pi}}{2\lambda} 
\int_{1}^{\infty} \frac{d\xi}{\xi^3}
\left[\frac{\lambda^2Q_s^2\xi^2+1}{\lambda^2\xi^2} 
     - \frac{1}{\lambda^2\xi^2} \right]
\nonumber \\
&=& 
\frac{\sqrt{\pi}}{2\lambda} ~Q_s^2 ~\frac{1}{2} ,
\end{eqnarray}
where we used the relation
\begin{equation}
\int_0^{\infty} dz E_1(z)= \int_0^{\infty} dz \int_z^{\infty} \frac{dt}{t} e^{-t}
= \int_0^{\infty} \frac{dt}{t} \int_0^{t} dz e^{-t}
= \int_0^{\infty} dt  e^{-t}=1
\end{equation}

We also derive the identity (62):
\begin{eqnarray}
 \int _{-\infty}^{\infty}d^2k_{1\perp} 
E_1(ak_{1\perp})E_1(bk_{1\perp})&=&
\frac{\pi}{2} \int _0^{\infty}dk_{1\perp}^2
\int_1^{\infty} \frac{dt}{t} e^{-ak_{1\perp}^2t}
\int_1^{\infty} \frac{ds}{s} e^{-bk_{1\perp}^2s}
\nonumber \\
&=&
\frac{\pi}{2} 
\int_1^{\infty} \frac{dt}{t} 
\int_1^{\infty} \frac{ds}{s} \frac{1}{at+bs}
\nonumber \\
&=&
\frac{\pi}{2} 
\int_1^{\infty} \frac{dt}{t} 
\left[ 
\frac{1}{at} \ln \left( \frac{s}{at+bs}\right) \right]_1^{\infty}
\nonumber \\
&=&
\frac{\pi}{2a} 
\int_1^{\infty} \frac{dt}{t^2} 
\left[ \ln(at+b)- \ln(b) \right]
\nonumber \\
&=&
\frac{\pi}{2b} 
\int_{a/b}^{\infty} \frac{dy}{y^2} \ln(y+1)
\nonumber \\
&=&
\frac{\pi}{2b} 
\left[
\ln \left( \frac{y}{y+1} \right) - \frac{1}{y} \ln(y+1)
\right]_{a/b}^{\infty}
\nonumber \\
&=&
\frac{\pi}{2} 
\left[
\frac{1}{b} \ln \left( \frac{a+b}{a} \right)
+\frac{1}{a} \ln \left( \frac{a+b}{b} \right)
\right]
\end{eqnarray}

\end{document}